\documentstyle[preprint,aps]{revtex}
\begin{document}
\title{The Real Significance of the Electromagnetic Potentials\thanks{%
Los Alamos National Laboratory e-Print Archive, quant-ph/9510004}}
\author{J\`un L\'\i u}
\address{Physics Department, Rensselaer Polytechnic Institute,
Troy, NY 12180-3590%
\thanks{%
{\it e-mail}: liuj3@rpi.edu.
{\it Web homepage}: http://www.rpi.edu/\symbol{126}liuj3}}
\date{October 5, 1995}
\maketitle

\tighten
\begin{abstract}
The importance of the potential is revealed in a newly discovered effect of
the potential.
This paper explore the same issue introduced in quant-ph/9506038
from several different aspects including electron optics and relativity.
Some people fail to recognize this effect due
to a wrong application of gauge invariance.
\end{abstract}

\section{Metaphysics: can we question the code we live by?}

This paper is about an effect which does not even have an appropriate name.
We can name it potential effect, effect of potential, effect of potential in
simply connected region, or, L\'\i u effect. I attach my name to the effect
for the reason that, As far as I know, I am the only one who advocate
this effect. Besides, there was no real obstacle to the discovery of this
effect any-time in the last fifty years.

We have existed in this world for only a limited time. We accept knowledge
at a limited speed. So our total knowledge is finite. Many of us
believe that the information in the world is infinite, or nearly infinite
when compared to what we have already received. We find rules, just like the
animals do, only much better. Beside our brain size, we
are doing much better largely because we
have a very good education system which can pass most of the knowledge from
generation to generation. A could-be experiment is to observe an individual
human who is isolated from all human knowledges.

The very personal knowledge we have is never accurate. What is the safe
distance between a lion and its prey? Neither party has an accurate number
for this. When the knowledge is recorded so as to be transmitted from
individual to individual, it must be coded on finite media for practical
reasons. At this point, knowledge becomes knowledges. The most basic code is
our language. There is obvious difference between the real knowledge and the
knowledges that piped through the code we use. We tend to regard the
knowledges as the knowledge most of the times, however, there are times when
the difference can not be ignored. The case I discuss here is one of such
cases.

We have other codes in science and technology beside language. Physics
formulas is an obvious example. Others are slightly difficult to explain in
a paragraph so I resort to an example. Newton said: ``F=ma, so ...'', while
he actually meant ``if F=ma, then we have ...''. In this case, F=ma is the
code, or gauge, to confine the way we record our knowledge of physics. It
turned out to be an efficient code, and thus remained ``correct'' for
centuries, until the discovery of relativity and quantum mechanics at the
beginning of this century. The problem is, people used to regard things like
F=ma as sacred, something to be taken for granted. We can record physics
with p=mv or G(a hypothetical quantity) = m(da/dt) as well as F=ma. F=ma is
chosen because it is more efficient to record the knowledge we know. Notice
that F=ma introduces two new quantities, F and m, not just one. Both F and m
are meaningless ( or mean something else, depend on the way you look at it,)
before the introduction of F=ma. Let me explain further for some readers.
The acceleration caused by gravity on most objects is (nearly) a constant on
the surface of the earth. The accelerations caused by a spring on most
objects are (nearly) constants. After the introduction of F=ma, we find it
possible to define m and F of most of the objects familiar to us as
constants. This in turn provides a concise language, or code, or gauge, to
record our knowledge of this world. There is nothing sacred about it. It is
just the way science is.

With the advantage of a coding system there comes disadvantage. People tend
to ignore cases when descriptions in accepted codes become complicated. This
is to some degree inevitable. We only have limited times. There are two
kinds of works in physics. The first kind is to spend time on recording the
world into knowledges under the current codes. The second kind is to spend
time on finding more efficient recording codes. Most of the great
discoveries belong to the second kind. However, there also seems to be a
third kind of works recently. It is about finding ``structurally
satisfactory'' recording codes witch is not necessarily efficient in
recording our world. They also spend enormous amount of time in recording
and converting existing data into these codes, because, the codes are indeed
``not necessarily efficient.'' There seems to be a misunderstanding between
the simplicity of the code, and its efficiency in recording our world.

The code relevant to this paper is electromagnetic gauge degree of freedom.
It is usually regarded as sacred and untouchable by some physicists. What is
this electromagnetic gauge degree of freedom? It is a degree of freedom
witch has no influence on observation. different ``values'' of it means
different ways of recording the same thing. To use a technical word, physics
is invariant under electromagnetic gauge degree of freedom. Conversion
between one way of recording to the other is called a gauge transformation.

So far so good. But what if we do see something which changes with respect
to the gauge choices? To some people, this is simply out of the scope of
interest. They put the factor to unity before a second thought. Therefore,
we do not even see enough explicitly expressed opinion about this. One
explicitly expressed opinion about this, which I would like to quote here,
is the opinion of Peshkin and Lipkin (1995) \cite{6}. They are both
well-known scientists in this curriculum. According to them, only
gauge-invariant quantities have physical meaning. Quantities that can not be
made gauge invariant have no physical meaning and therefore should be get
rid of. This opinion is quite common. Then they face the question of how to
get rid of these quantities. Different values of the gauge mean different
physical predictions. Which value should be chosen? The following opinions
of theirs is critical. They believe that the gauge can be, and should be
chosen in such a way that the quantity in question becomes not measurable. A
theoretical technique which is sometimes coined as ``gauge-away'' is
normally used to deal with these gauge-dependent quantities.

It is quite obvious that this ``gauge-away'' treatment is always less than
ideal. The philosophy behind such treatment is: ``we do not really know what
its value should be, so let us choose a value so as to keep the prediction
of theory as simple as possible.'' The reason that it is less than ideal is,
while simplicity is usually closer to the truth than complexity, simplicity
does not always equal to the truth in every segment of physics. Even if we
honor the simplification of physics as a whole, that still doesn't mean we
should always individually simplify every segment of physics to the extremes.

This is our discussion from a metaphysical point of view. We must have
unbiased opinions about all possibilities. If we made a choice of
convenience with the reason unknown, we should try to find the reason later.
If gauge invariance is true, yet we still see some quantities varying with
respect to gauge choices, then what are they? This question is obviously
reasonable. We used to say, there is reason behind everything. This
everything obviously also include the quantities which are gauged away. In
the next section, we shift to a physics perspective.

\section{Quantum mechanics as electron optics}

Aharonov-Bohm effect(AB effect) \cite{2} is a well-known effect in quantum
mechanics. It is so named because Aharonov and Bohm published a paper in1959
titled: ``Significance of Electromagnetic Potentials in the Quantum
Theory.'' The movement of an electron is influenced directly by potential,
in a region where there is no electromagnetic field at all!

Let me first briefly describe a typical setting of this effect. The basic
component is a long-enough solenoid, shielded from the travelling electron
by an impenetrable wall. The purpose of the shield is to ensure that the
electron only contact with the potential outside, not the magnetic field
inside the solenoid. This shielded solenoid is put in between the double
slits. The double slits would cast a typical sinusoidal interference pattern
on the screen. When a current is applied to the solenoid, the central
maximum would shift to one side. This effect, the AB effect, is caused by
potential only.

In this essay, I am going to talk about two surprises about this effect.
First, a small surprise: this AB effect was actually first discovered by
Ehrenberg and Siday in 1949. This is a well-known fact among the
specialists. Secondly, a big surprise: this AB effect is actually a
particular case of a more general ``potential effect.'' This was discovered
by me \cite{3} .

Everybody knows that quantum mechanics is not as perfect as special
relativity. Someone says: ``for quantum mechanics, Helium atom has one
electron too many.'' AB effect is another good example of this. We can not
predict AB effect from Schr\"odinger's equation of 1926. There was a long
time-span of thirty-three years between the two. Yet, thirty-six years have
passed after the publication of AB effect, and we still see little change in
quantum mechanics due to this. There are some works about solving the AB
problem from modeling in quantum mechanics. But the success is limited. The
two solenoids scattering problem, which should be a direct expansion of the
AB(one solenoid) scattering, is still under debate.

I must say a few words about electron optics in order to introduce the work
of Ehrenberg and Siday \cite{1} properly. Electron optics can be divided
into classical and quantum parts. In principle, everything should be
governed by quantum mechanics through wave-like treatment of the electron.
But in fact, the classical theory of electron optics still dominates.
Traditionally, quantum mechanics solves two kinds of problems: bound-state
problem and scattering problem. Approximation methods is almost always used.
Typical methods are: perturbation expansion and variation principle.
Electron optics, however, requires something more rigorous which standard
quantum mechanics can not provide.

Among many other physicists, Ehrenberg and Siday (1949) tried to solve this
problem. They defined electron refractive-index as a function of vector
potential. Near the end of their paper, they discussed ``a curious effect'',
which is exactly the AB effect: On the two sides of a magnetic flux, vector
potential has different values. This means different refractive index for
two geometrically equivalent paths. This difference in refractive index
would cause an observable phase-shift.

Now we are ready to get to the business of introducing potential effect \cite
{3}.

The potential effect can be introduced in many ways. Here I choose to do it
through electron optics. As we can see, wave electron-optics is, or should
be, a copy of quantum mechanics. The physical foundation must be the same.
In terms of electron optics, quantum mechanics is about how an electron
propagates in electromagnetic media. A bound-state of quantum mechanics is a
stationary propagation state, or, a standing-wave state, of electron optics.
In another words, a bound-state is a state that scatters into itself. What
about other interactions? First of all, quantum mechanics is primarily a
theory about electromagnetic interaction. Secondly, other interactions can
always be written in gauge formats similar to that of the electromagnetic
interaction. So for the moment we don't need to worry about the difference
too much.

Now let us have a version of wave electron optics:

\begin{enumerate}
\item  Electron propagates as a wave.

\item  The electromagnetic potentials behaves as a media of the propagating
electron. This media changes the local wavelength and frequency of the
propagating electron.
\end{enumerate}

I hope you can see the marvel of this theory. All quantum mechanics problems
can be solved this way. Computer simulation of wave-front can be used as a
handy method to solve practical problems this way.

Of course there are problems. Otherwise, it would be a popular method
already. The problem is gauge invariance. To solve propagating electron
problem this way, we need to know the ``refractive index'' of each point
exactly. It seems that a gauge transformation would cause change of the
media. So here we encounter another quantity which is gauge dependent.

The issue of gauge-dependent quantities is already discussed in section one.
The prescription of ``gauge away'' obviously doesn't work here, because that
would result in plane-wave-only solution for all quantum problems of
electron! This is in the blind point of some physicists. It is thus ignored
as a possible method due to this impasse of gauge dependence. It is
interesting to learn that, most basic problems in electron optics are solved
using classical electromagnetics, in which the electron is represented as a
particle.

Now let us revisit the AB effect. In AB effect, the electron goes through
two different paths. Normal gauge transformation does not change the optical
path-length difference between the two paths. Therefore, AB effect has no
trouble with gauge invariance.

Potential effect \cite{3} is different from AB effect in the sense that it
means observability of the effect of potential in any regions, not just
multiply connected regions. Thus, AB effect is just a special kind of
potential effect \cite{3}. The simply connected region is obviously the best
example and testing ground of this effect. The notion of the potential
effect also fixes the ``refractive index'' uncertainty in the
above-mentioned problem of electron optics. Thus, the existence of the
potential effect also means a fundamental change in our views of quantum
mechanics.

How to observe the effect of potential in a simply connected region? It is
actually very simple. The only additional instrument needed is an elongated
toroidal solenoid which can provide a constant vector-potential region
inside. When an electron travels inside, the local wave-length would change,
and this change is measurable by any interference experiment. Wave-length is
always the first observable quantity in any experiment that shows the
wave-like nature of the electron. The experiment can be double slit, single
slit or even holography.

People haven't seen fundamental changes in physics for a long time.
Therefore, it is quite understandable that they doubt my discovery when they
see it for the first time. It has been three and a half years since my first
attempt to publish the result. The following two opinions are typical: (1)
``Of course the effect exists and there is no news here. It is a simple
variation of AB effect.'' (2) ``The effect can not possibly exist because it
violates the principle of gauge invariance.'' We should notice that point
one contradicts point two. So it is physically illogical for one to
say:``either (1) is true, or (2) is true.'' By very simple logic we can
already reject one of them.

Of course both points are wrong. The effect exists and it is revolutionary.
In the follow I discuss this problem from the perspective of Lorentz
invariance. After that, I would like to direct you to my paper \cite{3}.

The theory of special relativity states that time and space can be combined
to be called as space-time. They are exchangeable. The description of a
wave-front propagating with respect to time is somehow non-relativistic. The
relativistic description of this is, a ``stationary'' state existing in
space-time.

The point is, phase comparison can be conducted between any two space-time
points, rather than just between two spatial paths. This is another
perspective to look at the nature of this effect of potential. It can be
seen as a generalization of AB effect. An electron coheres with itself in
time, and remembers where it comes from. When an electron travels into a
region with different potential values, its wave-length and frequency change
accordingly. This change in time is for real, as real as the phase
difference between two different paths created by potential in multiply
connected regions.

Most experts in this field do not believe this effect can exist (in simply
connected region) because they think there is no reference for the electron
to compare its phase with. And they therefore believe that gauge
transformation means that the phase-factor is uncertain. But how they
establish equality between an uncertain, gauge-dependent phase factor and a
zero phase factor is something beyond my understanding. As mentioned above,
I believe the existence of this effect because, among other reasons, time
coherence is just as reasonable as spatial coherence. This is a consequence
of Lorentz invariance.

AB effect can be easily derived from potential effect in the simply
connected region. We simply regard the double path as the combination of two
simply connected paths. The phase shift in the two paths will be different.
The difference between the two is the AB phase shift.

The existence of this effect of potential means a brand-new version of
quantum theory. This effect fixes the uncertainty in above-mentioned
electron optics. So logically, all quantum problems can be defined and
solved in this way. We no longer logically need any eigenstate equations.
All we need is a proper handling of the quantity p-eA while treating
electron as a propagating wave. This is the real significance of the
electromagnetic potentials.

\section{Classical and quantum electromagnetics}

When a particle is generated, its 4-momentum
$$
i\hbar \partial _\mu +qA_\mu
$$
should have unique expectation. The natural way of doing this, is define $%
qA_\mu $ to be zero at this point. These conditions pin-down the gauge
degree of freedom. The potential thus becomes uniquely defined for this
particular particle.

In practical calculation, we do not always have to start from a point
source. We can also start from any intermediate wave-front. But similar
conditions apply and the gauge is fixed. This is a quantum phenomena. The
gauge enters the propagation through the phase factor of the wave-function
only. The fact that phase is a constant on a wave-front means that the gauge
can be treated as the phase factor. If a quantity has absolutely no impact
on measurement, than that quantity should be recycled.

The {\em potential effect} \cite{3} is a reflection of this significance of
potential. The effect is detectable not just in some special multiply
connected region, like the AB effect, but also in everywhere when the change
of 4-wavelength becomes measurable. A typical and simplest example is its
effect in a simply connected region. And a typical experimental realization
of this example is any wavelength sensitive electron experiment inside an
elongated toroidal solenoid. Since the gauge is fixed, the potential value
inside the toroidal solenoid is directly related to the wave-vector {\bf k}
of the electron when it is inside the toroidal solenoid. If we use the
traditional double-slit setting, the change of current in the toroidal
solenoid would cause change of fringe-spacing on screen.

We can in principle use this fixed gauge to define reflective index. After
that, computer simulation becomes very handy. It could become the best
option of solving complicated problems because the method is numerically not
sensitive to the complexity.

\subsection{Number of variables}

We used to describe electromagnetic interaction on the classical level in
terms of the fields {\bf E} and {\bf B}. We should notice that all of the
classical electrodynamics \cite[for example]{7} can be formulated in terms
of the potentials $\Phi $ and {\bf A}, instead of {\bf E} and {\bf B}. We
simply replace the fields using%
$$
F_{\mu \nu }=\frac{\partial A_\mu }{\partial x_\nu }-\frac{\partial A_\nu }{%
\partial x_\mu }.
$$
We can see that, on the classical level, the potential is a simpler code
than the fields for the description of electromagnetic interaction. This is
obvious in the sense that the potential has a total of four components while
the fields have a total of six components.

The potential description on the classical level is almost perfect save one
thing: the gauge invariance. Due to the gauge invariance, there are only
three independent physical components on the classical level. This fourth
variable, the gauge, is actually connected to the wave-function of quantum
mechanics. This gives an un-separable coupling:
\begin{equation}
\label{gauge}qA_\mu \rightarrow qA_\mu +\partial _\mu G,\;\text{and }\;\psi
\rightarrow \psi \times exp\left( \frac i\hbar G\right) .
\end{equation}
In some sense, this gauge is no longer {\em invariant} in the quantum
region. Of course, we can always construct invariant quantities. In this
case, it is
$$
\left( qA_\mu +i\hbar \partial _\mu \right) \psi .
$$
This becomes known as the canonical momentum, or, the minimum coupling. One
may argue that we have the minimum coupling first then we construct \ref
{gauge} from it. How it come to be is not important. The important thing is,
this minimum coupling have been indirectly proved through the very accurate
experiments of QED.

Now let us count number of variables again. Since the wave-function $\psi $
is complex, it counts as two variables. One of them is coupled through the
gauge G. So, the number of total independent variables is $4+2-1=5$. This 5
may sound strange to some people. We count both the potential which provides
influence, and the quantum representation of the traveling quantum particle.
The reason they are counted together is because they are coupled together.

\subsection{Current-current interaction}

On a more fundamental level, we must understand interaction on the basis of
direct particle-particle interaction. In the case of minimum coupling, The
potential is an indirect representation of the currents - a lot of moving
particles. We now express $A_\mu $ in terms of 4-velocity of the
potential-providing particles.

The notion of a conserved four canonical momentum is the back bone of the
effect. We can re-express it as:

$$
const.=p_\mu \left( x\right) -qA_\mu \left( x\right)
$$
$$
{\bf =\hbar }k_\mu \left( x\right) -q\sum_iq_i\int d\tau D_r\left(
x-r_i\left( \tau \right) \right) V_{i,\mu }\left( \tau \right) ,
$$
where $V_\mu $ is the four-velocity, $D_r\left( x-r_i\left( \tau \right)
\right) $ is the retarded Green function $\cite[p. 654]{7}$, summation is
over all particles except the one which is represented as ${\bf \hbar }k_\mu
\left( x\right) $. In the case of electron-only, we have%
$$
k_\mu \left( x\right) =k_{0,\mu }+\alpha \sum_i\int d\tau D_r\left(
x-r_i\left( \tau \right) \right) V_{i,\mu }\left( \tau \right) .
$$
This interpretation means the influence of the electron is not bounded. It
can affect other particles through a propagator. That propagator determines
all features of classical and quantum theory of electromagnetics. We expect
the result of higher order effects comes from radiative corrections of both
the potential-source and the selected particle.

The particles that provide the potential and the particle that receives the
potential are not represented symmetrically here. If we consider the quantum
uncertainty of the potential provider, we need another degree of freedom.

\section{Eikonal approximation}

The complex wave-function is defined as%
$$
\psi \left( {\bf r},t\right) =a\left( {\bf r}\right) exp\left( \frac i\hbar
\left( S\left( {\bf r}\right) -Et\right) \right) .
$$
Notice that in this definition, energy eigenstate is automatically assumed,
while in relativistic case this is not correct. Also, non-relativistic
quantum mechanics
\begin{equation}
\frac 1{2m}\left( -i\hbar \nabla +e{\bf A}\right) ^2\psi -e\Phi \psi =i\hbar
\frac{\partial \psi }{\partial t}
\end{equation}
is usually adopted. After separation of real and virtual parts, the result
is two real equations for $a\left( {\bf r}\right) $ and $S\left( {\bf r}%
\right) $:
\begin{equation}
\label{1}\left( \nabla S+e{\bf A}\right) ^2=2me\Phi +2mE+\hbar ^2\frac{%
\nabla ^2a}a,
\end{equation}
and
\begin{equation}
\label{2}\nabla \cdot \left[ a^2\left( \nabla S+e{\bf A}\right) \right] =0.
\end{equation}
These two equations are well-known, and can be interpreted as
Hamilton-Jacobi equation and current-conservation equation. The eikonal
approximation is good when the last term of \ref{1} is relatively small. To
put in word, the change of amplitude should be smooth enough. After the
approximation,
\begin{equation}
\label{eikonal}\nabla S=\widehat{{\bf t}}\left( {\bf r}\right) \sqrt{2me\Phi
+2mE}-e{\bf A.}
\end{equation}
where $\widehat{{\bf t}}\left( {\bf r}\right) $ is unit vector which can be
determined in principle, after the solution of Schr\"odinger equation has
been found.$\cite{5}$ If we have $\widehat{{\bf t}}\left( {\bf r}\right) $
beforehand, we can integrate it in the curl-free case.

$\widehat{{\bf t}}\left( {\bf r}\right) $ is usually assumed to be the
classical trajectory of the electron. This means, in the curl-free case, $%
\widehat{{\bf t}}\left( {\bf r}\right) =const.$ We can also do the same for
$%
t_\mu \left( x\right) $. But this is obviously an approximation. In cases
like diffraction around sharp edges, $\widehat{{\bf t}}\left( {\bf r}\right)
$ is obviously not constant. In this sense, the eikonal approximation is
semiclassical. We expect it to be valid in the paraxial region in electron
optics.

\subsection{Lorentz invariance}

Lorentz invariance requires equal treatments of all 4-variables. The mass of
the electron is a scalar, but energy is not a scalar. Energy eigenstate is
no longer a preferred description. We must in principle allow frequency to
change in space-time. This covariance argument leads to the following
generalization of eikonal approximation.

The derivation should start from the Klein-Gordon equation if we do not
consider spin. Also, the consideration of gauge invariance is taken into
account. G in the follow {\em is} the gauge.
\begin{equation}
\left( i\hbar \partial ^\mu +qA^\mu \right) \left( i\hbar \partial _\mu
+qA_\mu \right) \psi =m^2\psi ,
\end{equation}
where q is the charge. After letting
\begin{equation}
\psi \left( {\bf r},t\right) =a\left( {\bf r},t\right) exp\left( \frac i\hbar
G\left( {\bf r},t\right) \right) ,
\end{equation}
where a and G are real. After separation of real and virtual parts, we have
\begin{equation}
\left( qA^\mu -\partial ^\mu G\right) \left( qA_\mu -\partial _\mu G\right)
=m^2+\hbar ^2\frac{\partial ^\mu \partial _\mu a}a
\end{equation}
and
\begin{equation}
\partial ^\mu \left[ a^2\left( qA_\mu -\partial _\mu G\right) \right] =0.
\end{equation}
These two are the relativistic version of basic equations. The eikonal
approximation can be introduced as:
\begin{equation}
qA_\mu \left( x\right) -\partial _\mu G\left( x\right) =t_\mu \left(
x\right) m.
\end{equation}
The m here is the electron rest-mass. This can be called covariant eikonal
approximation. $t_\mu \left( x\right) $ can also be identified as classical
trajectory.

This is however not the one to be compared with the non-relativistic eikonal
approximation. The energy appears in \ref{1} and \ref{eikonal} is%
$$
E=mc^2\left( \frac 1{\sqrt{1-{\bf v}^2/c^2}}-1\right) .
$$
The another choice of approximation using a three-dimensional ${\bf t}\left(
x\right) $ is:
\begin{equation}
\nabla G-q{\bf A}=\widehat{{\bf t}}\left( x\right) \sqrt{\left( \frac{%
\partial G}{\partial t}-q\Phi \right) ^2-m^2}.
\end{equation}
This also means that we see it as an energy eigenstate. Just like we can
always have the third component (z) of angular momentum $J_z$ determined
simultaneously with $J^2$, we can have energy eigenstate together with the
``mass eigenstate''. This way, it is comparable to the non-relativistic
eikonal approximation. However, we should bear in mind that this is a
choice, not a must.

\section{Application}

The discussion here is mostly conceptual. The effect is so straightforward
that we need very little mathematics. The application of the effect is not
discussed here because they are consequences of this prediction and are
logically connected. The immediate application is the extension of AB
effect. We can apply this effect of potential where AB effect is observed.
The second application is in electron optics \cite{4} . Beside these
applications, we view this theory as a new way of solving general quantum
problems. It reveals the real physical significance of the electromagnetic
potential.

\end{document}